# MINUET: A METHOD TO SOLVE SUDOKU PUZZLES BY HAND

Carlos F. Daganzo
Professor of the Graduate School
University of California, Berkeley, CA 94707
(December 24, 2018)

## ABSTRACT

This paper presents a systematic method to solve difficult 9x9 Sudoku puzzles by hand. While computer algorithms exist to solve these puzzles, these algorithms are not good for human's to use because they involve too many steps and require too much memory. For humans, all one can find in the literature are individual tricks, which used together in ad hoc ways can be used to solve some puzzles—but not all. To the author's knowledge, a systematic procedure made up of well-defined steps that can be carried out by hand and solve all puzzles has not been devised. This paper proposes one such technique—the minuet method. It is based on a new system of markings and a new way of simplifying the puzzles that can be easily carried out by hand—or by computer.  The author has solved hundreds of puzzles of the most difficult kind ("evil" in Sudoku's parlance) and never found one that could not be solved. The average time to solve one of these puzzles is slightly over 1 hour. It is conjectured that this method can solve all well-posed 9×9 puzzles.

The method's distinguishing feature is a *minuet* strategy that is applied when the puzzle cannot be further simplified with basic tricks. The strategy consists in concurrently developing two potential solutions, sometimes alone and sometimes in concert as if they were dancing a minuet. The procedure is recursive. Each stage involves the following: (i) *choose dancers*: identify a binary choice (say, a cell that can be solved by only two numbers or else a number that must solve one of two cells), one of which must be "true" and the other "false"; (ii) *dance alone* I: develop a solution as completely as possibly using one of the choices as a starting point, and using a special set of markings to record the results; (iii) *dance alone* II: do the same with the other choice using a different set of markings; and then, (iv) *dance together*: consider both sets of markings together to extract further simplifications. These do not emerge if each of the choices is considered individually. This is why the method is powerful. Steps (ii)-(iii) are repeated as in a minuet until the puzzle can no longer be simplified. If this occurs, which is rare, the minuet is repeated with another pair of dancers until a solution is found.

## 1. INTRODUCTION AND OVERVIEW

Sudoku is a combinatorial game that has garnered much attention with the public because systematic methods to solve these puzzles by hand apparently have not been found. This, after all is what makes them fun. Sudoku has also attracted the attention of mathematicians because as shown in [1] puzzles are in a class of combinatorial problems known as "NP-complete"; see [2] for a definition.

Besides complexity, mathematicians have explored other properties of the problem, such as enumerating the number of solutions [3] and identifying minimum number of clues under which a unique solution exists [4] (now known to be 17).

More relevant for our purposes, mathematicians have also explored computer solution algorithms of different types. These include brute-force searches, which are not very interesting especially for humans, and more efficient approaches. Among the latter are combinatorial optimization approaches of the type that can be solved with systematic search algorithms. Reference [5] is an overview of exhaustive methods in this discipline. Applications to Sudoku include branch-and-bound [6] and stochastic metaheuristics such as simulated annealing and its variants [7]. Another efficient approach casts the problem as a constraint problem [8][9]. Unfortunately, all the computer algorithms seen by this author exploit features of computers that are not shared by humans, such as rapidity of calculation and a prodigious memory that can be erased and rewritten at will. While this is good to create puzzles online and produce solvers that come up with solutions at the push of a button, it is of no use to the Sudoku enthusiast who tries to solve them with pencil and paper.

The quest here is to find an algorithm that can be implemented not just in a computer but also be used by humans to solve any well-posed puzzle in a reasonable time, say in less than two hours, with only pencil and paper. The author came up empty when researching the existence of such a method. An Internet search revealed a plethora of tricks to simplify puzzles (way too long to list here) but no systematic approach. Closest to the goal is an online post [10] with a sequence of tricks (some quite clever such as the "x-wing pattern"; "swordfish pattern", "solving with colors" and "forcing chains") that is claimed to solve some many puzzles but is known to fail on some [10].

Rather than a set of tricks, this paper describes a *method* that is conjectured to solve all well-posed Sudoku puzzles; i.e., puzzles that have a unique solution. The method is systematic in that it could be programmed for a computer (with a long program) but is designed for humans. The method uses just a few tricks. These include some that are well-established such as hunting for "singles" and "doubles", and the writing of "candidates" for later elimination (see e.g., [10] for explanations, or the description in this paper) as well as some new tricks. The proposed method does not use other established tricks, such as the above-mentioned "x-wing" and "swordfish".

The method's distinguishing features are its "minuet" step and the set of markings used to implement it. The author has solved hundreds of Sudoku's of the most difficult kind, and the method has never failed. This is the basis for the conjecture. The only tools needed for human application are: pencil, paper and some form of permanent marking that we call "ink". The method involves lots of erasing, and requires much attention to prevent errors. If the method is programmed for computer, counterexamples to the conjecture and/or additional properties could emerge.

The method has two phases: A first phase where the Sudoku grid is populated with possible numbers, which we call "candidates", and a second phase where these candidates are eliminated from the grid until only one number remains in each cell.

Each phase consists of two steps that achieve different things. The description below assumes that these steps are performed iteratively in a particular sequence. Although this works well, experience shows that the solution time (by humans) can sometimes be reduced if at opportune times one deviates from the suggested sequence and uses elements of a step when performing another. In the interest of brevity, acceleration tricks of this type will not be discussed—readers may have fun discovering them on their own.

The paper starts with some preliminaries, including terminology. It then presents the solution method.

## 2. BASICS

This section describes both basic ideas used by the method and the terminology that will be used to describe it. Some of the terminology has been made to coincide with that used in [10]—but not all since some of the concepts are new. The following is a list of terms, with comments.

1. *Sudoku grid*. A 9x9 grid of empty cells. As every Sudoku player knows, this grid has nine rows, nine columns, and nine 3x3 squares bounded by heavy lines.

2. *Structure*: Any of these rows, columns, or 3x3 squares.

3. *Basic Sudoku Rule*: Place one and only one number from 1 through 9 in each cell so that each structure contains exactly one instance of each number.

Consideration shows that the Basic Sudoku Rule is equivalent to the following two sub-rules:

*Sub-rule* 3a: A number cannot appear more than once in each row, column, or square.
*Sub-rule* 3b: Each number must appear at least once in each structure.

4. *Conflict*: A violation of Sudoku Sub-rule 3a; that is, more than one instance of the same number in a structure.

5. *Candidate (for a cell)*: A number that could go in a cell without violating Sub-rule 3a, that is, without a "Sudoku conflict." Until a Sudoku is solved, numerous cells can have more than one candidate.

6. *Unsolved cell*: a cell whose number has not yet been identified; as opposed to a *solved cell*. Unsolved cells have more than one candidate.

7. *Single*: A number that has been identified as solving a cell and the cell containing the number.

Singles are very important as they form part of the solution. They come in two flavors: "naked" and "hidden".

> 8. *Naked single*: A number that is the only candidate in a cell, and the cell in question. Since said number is the only feasible solution for the cell, it solves the cell as per the Basic Sudoku Rule. The single is said to be "naked" because no other candidates surround it to hide it from view.

> 9. *Hidden single*: A number that appears as a candidate in only one cell of a structure, and the cell in question. There may be other candidates in the cell. This is a single because the number solves the cell as per Sudoku Sub-rule 3b. The single is hidden because it may be obscured by the other candidates in its cell.

Note there is a duality between naked and hidden singles. One finds naked singles by looking at an individual cell of a particular structure and seeing if there is a unique number that can go into it. One finds hidden singles by doing the opposite: Looking at a number and seeing if there is a unique cell of a structure that can contain the number.

Almost as important as singles are doubles.

> 10. *Double*: A pair of numbers that have been identified as solving two cells (in an unspecified order) and the two cells that contain them.

Like singles, doubles can also be naked or hidden, and the duality persists. In the naked case, two cells of a structure contain only two candidates, which are the same in both cells; and in the hidden case a pair of numbers appears in only two cells of a structure. More specifically, we have:

> 11. *Naked double*: A pair of numbers that are the only candidates in two cells of the same structure, and the two cells that contain them. No other candidates appear in these cells. (The numbers in question can also appear in other cells of the structure, albeit accompanied by other candidates.) The pair of numbers and the two cells that contain them exclusively form a double because the Basic Sudoku Rule requires each of the two cells to be solved by a different number. Since only two numbers are available the cells are solved by placing one number in each cell. Thus, the pair solves the cells.

> 12. *Hidden double*: A pair of numbers that appear as candidates in only two cells of a structure, and the two cells that contain them. As in the case of hidden singles, the cells can contain other candidates. However, since our two cells are the only locations of the structure where the two numbers can appear, and the numbers must appear at least once as per Sudoku Sub-rule 3b, it follows that the two numbers solve the cells and form a double.

The following is an additional important concept still regarding doubles that will be used repeatedly.

> 13. *Half double*: a number that is a candidate in only two cells of a structure, and the two cells that contain it.

Half doubles are important because they are quite common, easy to spot and they help simplify the puzzle. In particular, note the following fairly obvious corollary that can be used as you solve the puzzle:

*Corollary* 13a: *Two half doubles in the same two cells are a hidden double.*

The ideas of singles and doubles can be extended to larger groupings. For example, we also have the following.

14. *Triple*: A trio of numbers that have been identified as solving three cells (in an unspecified order) and the three cells that contain them.

15. *Naked triple*: A trio of numbers that are the only candidates in three cells of the same structure, and the three cells that contain them. For example if the only candidates of three cells in a structure are {3, 6}, {3, 7} and {3, 6, 7} then these cells and the numbers {3, 6, 7} are a naked triple.

16. *Hidden triple*: A trio of numbers that appear as candidates in only three cells of a structure, and the three cells that contain them.

Hidden triples can sometimes be identified with the help of half doubles as per the following fairly obvious corollary:

*Corollary* 16a: Three half doubles that occur in only three cells of a structure are a hidden triple.

Although one could also define and look for quadruples or higher groupings, the proposed method shall not use them. Looking for them turns out to be unnecessary. There are two final important concepts:

17. *Blocking*: The act of preventing a number from appearing in a cell or a structure by way of a conflict. Example: the number in a solved cell blocks itself from any unsolved cell in the same row (or column or square). This follows from basic Sudoku Sub-rule 3a,

18. *Unavailable cell* (*to a number*): A cell that is blocked to the number.

Singles, half doubles, doubles and triples enforce blocking actions that help simplify puzzles. Here are some useful blocking rules:

*19. Rule for Singles*: The number of a single appears alone in its cell, solving it. Therefore, the number is blocked from the three structures that contain said cell.

20. *Rule for Half Doubles*: The number in a half double blocks said number from all the structures containing the half double. Why? Since the number must appear in one of these two cells (as per Sudoku Sub-rule 3b) the number cannot appear elsewhere in any structure that contains the half double (as per Sub-rule 3a).

21. *Rule for Doubles--or Triples*: The numbers in a double--or triple--block said numbers from any structures that contain the double or triple (for the same reason as happened for half doubles). Furthermore, the cells in the double (or triple) are unavailable to all other numbers. Why? Because the two—or three—cells must be populated by the two—or three—numbers (in an unspecified order) as per the definition of double-or triple; therefore, other numbers are blocked from these cells by the Basic Sudoku Rule.

22. *Passive Rule for Half Doubles and Doubles*: If a cell in a half double (or double) becomes unavailable to the number of the half double (or one of the numbers of the double), then the other cell and the number that was blocked from the first cell become a hidden single. <u>Example</u>: This occurs if a number other than the one of a half double turns out to solve one of the half double's cells; and also if the number of the half double appears in a structure that overlaps with one of the cells of the half double.

The steps of the proposed method build on Rules 19-22, and on Corollary 13a. Let us see how.

## 3. METHOD

The method is iterative and requires erasures. Therefore, pencil will be used to write the cells' candidates, as well as other markings that may have to be erased. Ink will only be used to write permanent results; e.g., the number that solves a cell. The problem's data is also displayed in ink.

Let us now look into the method's two phases. Recall that in Phase I the Sudoku grid is to be populated with candidates, and in Phase II these candidates are systematically erased until only one number remains in each cell.

### 3.1. Phase I: Filling in the candidates

The goal here is identifying as few candidates for each cell of the grid as possible, *but without missing any*. This is done in two steps: In Step 1, numbers will look for cells, and in Step 2 cells will look for numbers.

<u>Step 1: Look for hidden treasures in the squares:</u>[1] The idea is to unveil as many hidden singles and half doubles in each square as possible; and also hidden doubles. It's good to be thorough, but not essential. Found singles will be written in ink and half doubles in pencil. For the latter we will use small digits confined to the upper left corner of the cells. In addition, if Corollary 13a reveals hidden doubles they will also be written in pencil. For these we shall use big digits in the center of the cells. Big digits signify candidates; i.e., that the cell in question is not available to other numbers. The small digits in the upper left corners do not block the cells to other numbers. The basics of Step 1 are now described. (Non-essential tricks are relegated to footnotes that can be skipped without loss of continuity.)

---

[1] The procedure about to be described can also be applied to rows and columns. This is not recommended, however, because the few additional treasures that are sometimes unveiled by considering all three structures are quickly found in Step 3.1.

We first do a first pass, which consists in considering one number at a time, and for each square that does not contain the number in ink doing the following:

Step 1.1: *Identify hidden singles for the considered number and square.* Look for all the cells in the square that are available to the number (i.e., not filled with a solution or other candidates) and determine in which of these the number can go without a Sudoku conflict. If only one cell remains available, then the number in this cell is a hidden single. This is true because the number cannot go anywhere else in the square. So, pat yourself on the back and write it in ink. The cell is now filled and no longer available to other numbers. With practice the cells that remain available can be identified very, very quickly.[2,3]

Step 1.2: *Identify half doubles for the considered number and square.* If you determine with the procedure of Step 1.1 that exactly two cells remain available, write the considered number in pencil on the upper left corner of these cells (really small). These cells (together with the square and number) are a half double for the square in question. The small penciled numbers are a useful way of remembering these blockages, so that later in the puzzle you can use Rules 20 and 22.

Step 1.3: *Identify hidden doubles for the considered number and square.* On occasion a half double will occupy the same two cells as a previously identified half double. Then, as per Corollary 13a, the two cells and two numbers become a hidden double. Since the two numbers are then the only possible candidates for each of our two cells you can write the two numbers (big) in pencil and use this information as you proceed.[4] Recall that Rule 21 implies:[5] (i) that these cells become unavailable to all other numbers; and (ii) that the two numbers are blocked from the remainder of the square, and if the cells happen to be in the same row (or column) also from the remainder of this row (or column).

It bears repeating that to be as efficient as possible, when you analyze a square-number combination you should consider all the cell blockages and unavailabilities implied not just by the original data but also by all the singles, half singles and hidden doubles (and possibly triples) you have identified from previous squares and numbers.

---

[2] One approach consists in determining which of the six rows and columns that cover the square contain the number outside the square either as a single (in ink), or else as a hidden double or double (in pencil). Since both of these occurrences produce a conflict (as per Rules 19 and 20) you can then eliminate from consideration all the cells in the square that are covered by these rows and columns, as well as those cells that are not available to begin with. All other cells remain available.

[3] Additional tricks can be used. For example, if you discover a single in a cell of a half double of a previously considered number, then you can apply Rule 22 and write a second single for the previously considered number (in ink) in the cell that remains available to it. Similarly, if during a second pass of Step 1 you identify a single for one of the cells of a double you can write a second single for the other cell of the double. You can also invoke Rule 22 if you notice that the number of a half double (or one of the numbers of a double) appears either as a single, or twice as a double, in a row or column that covers on of the cells of the half double (or double). In this case too, you can write the number you are considering as a single (in ink) in the other cell of the half double (or double).

[4] Optionally, you can also look for hidden triples using Corollary 16a.

[5] The implication also holds for triples.

The first pass ends when you have explored all the number-square combinations. Since you will have filled more than a few cells with singles, hidden singles and doubles, you should then perform another pass as this will often yield additional singles, hidden singles and doubles (and on occasion triples). If you find some, you should perform yet another pass. Step 1 ends when you cannot find more hidden singles, half doubles and hidden doubles. This usually takes two or three passes.

At this stage some cells will be solved (and filled in ink) and others will contain two candidates in big penciled numbers--or perhaps three if you have looked for triples and found some. The rest will be empty of candidates, although they may have small numbers corresponding to half doubles. The next step is filling these voids.

Step 2: Fill all cells with candidates. The objective of this step is populating all the empty cells with candidates. It is very important not to miss any candidates. To do this, you should consider one empty cell at a time, and then write in pencil all the numbers that are not blocked from being in the cell. Remember that a number is blocked if it appears in any structure containing the cell either as a single (a solution in ink), a double, a triple, or as a half double. Recall that the latter are indicated by small pairs of numbers *in cells belonging to the same square*.

If a cell turns out to have only one candidate, it is a naked single. Therefore it solves the cell and you can write it in ink. Whether you find any naked singles or not, by the end of this step all the cells should be filled either in ink or in pencil.

From now on, the procedure systematically prunes the penciled in candidates until only one remains in each cell.

### 3.1. Phase II: Pruning the candidates

This also consists of two steps. The first of these (Step 3) is the most basic.

Step 3: Basic cleanup: The idea is to scan every structure (27 in all) for naked and hidden singles, doubles and triples (higher groupings are not necessary) and to clean up after each find. The cleanup is done by exploiting the new blockages and cell unavailabilities that each find implies. This step is so fundamental it will be used as an element of Step 4. The mechanics are now described in detail.

The following three sub-steps should be applied to each structure:

*Step* 3.1*: Find singles.* Naked singles are easy because they stand out. They are in cells that contain a single candidate—the naked single. You can write them in ink. Hidden singles are a little trickier to find. Those associated with squares should have been identified at the end of Step 1, but some may remain if you have not been thorough. More likely is that you will find some associated with rows and columns, which were not considered in Step 1. To identify singles for the structure under consideration (be it a square, row or column) look for numbers that appear only once as candidates in the structure. These are your hidden singles. You can also write them in ink. After identifying a single of either type you should then "clean up" with Rule 19: (i) Erase any other candidates in the cell of the single; and (ii) erase all candidates blocked by the single in the three structures that contain it.

*Step* 3.2: *Find doubles.* Look for naked and hidden doubles in the structure under consideration. It should perhaps be intuitive that this is only necessary if the structure in question has four or more unsolved cells.[6] Naked doubles are easily found: look for cells in the structure with only two candidates that match. Hidden doubles are found systematically by recording all the numbers that appear as candidates exactly twice in the structure[7] and then checking if any two happen to be in the same two cells. Once a double has been identified (of either type) you can use Rule 21 to execute the cleanup: (i) erase any candidates that are not part of the double from the cells that contain the double; and (ii) erase all candidates that the double blocks in the structure (or structures) that contain the double. [Note: There is always one structure containing the double, but there is an additional structure if the two cells of the double happen to be in the same row or column of a square.]

*Step* 3.3: *Find triples.* Triples need to be searched only if the structure under consideration has 6 or more unfilled cells. Like, naked pairs, naked triples stand out, albeit not as clearly. They consist of three cells with only 2 or 3 candidates, all members of a common set of three numbers; e.g., the cell contents could be {5, 6}, {6, 8} and {5, 8}, which are in the set {5, 6, 8}. Hidden triples are trickier. They are systematically found by identifying all the candidates that appear either two or three times in the structure (you can write them in pencil on the side of the puzzle) and then checking if any three happen to appear in the same three cells of the structure—and in no other cells. If a triple of either kind is identified you can again invoke Rule 21 and clean up; i.e.: (i) erase any candidates that are not part of the triple from the three cells that contain the triple, and (ii) erase all candidates that the triple blocks in the structure(s) that contain the triple.

*Quadruples*: They can only occur in structures with 8 or more unsolved cells and are extremely rare. Don't bother looking for them.

Once you have considered all the structures, you may have eliminated many candidates. As a result, more singles, doubles and triples may have emerged. For this reason, you should repeat Step 3 until you cannot find any more singles, doubles or triples. With practice, you can do this fairly rapidly; for example by learning to identify which structures and numbers are likely to yield simplifications after a cleanup.

Step 3 is very powerful. In fact, most easy and intermediate level Sudoku puzzles are often solved after completing Step 3. But the more difficult ones require one more step.

---

[6] Obviously, if a structure with three unsolved cells had a double then the other unsolved cell would have to be a hidden single, which would already have been identified when looking for singles.

[7] These are half doubles. If in a square, the numbers are recorded in their own cells as in Step 1.2. If in a row or column, the number should be written on the margin of the corresponding row or column. These markings will be used in Step 4.

Step 4: The Minuet: The idea here is to identify a pair of potential solutions (the dancers) that will help us trim the puzzle when they dance both alone and together. The two solutions emerge from any unresolved binary choice; i.e., where we know that one of the choices is "true" and the other "false", but we don't know which is which. An example of a binary choice is a "starter cell" that contains only two candidates—since one and only one of the candidates must be the correct choice for the cell. Another example is a "starter half double"—since one and only one of the cells of the half double is solved by the number of the half double. The items chosen are called "starters" because they are used to initiate the two dancing solutions.

Considering the two potential solutions (dancing) together is useful because candidates that cannot be part of either solution can be eliminated. This is more fruitful than considering each potential solution alone. Dancing eliminations should be followed by a basic cleanup (Step 3) for the whole puzzle, and then by a return to the dance (Step 4) to eliminate more candidates. This alternating process between Steps 3 and 4 can be continued until one of the potential solutions emerges as the correct one. The specifics are as follows. An important detail will be a system of markings that will allow us to track both dancers together with no clutter and little erasing.

To begin, identify a starter item (e.g., a cell with only two candidates or a half double).[8] Enclose one of the two starting choices in a circle and the other in a square. Then, considering one choice at a time, use Step 3 to eliminate candidates, and in this way develop as much of each solution as you can (this is when the solutions are dancing alone).[9] The only difference now is that instead of erasing conflicted candidates we merely highlight them. In the description that follows we do this by using small marks for the *retained* candidates, although marks could instead be used for those that are eliminated.

The author uses overdots for the circle solution and underdots for the square solution. To avoid too many markings, the author does not use overdots (underdots) when all the candidates are retained in the circle (square) solution. This code allows one to perform Step 3 for each of the solutions separately (dancing alone), including the identification of singles, half doubles and doubles. For additional clarity, when only one dot remains either above or below a number (i.e. the number is a single in the corresponding solution) the number is enclosed in either a circle or square.

The power of the minuet method rests in the fact that the patterns of markings for the two solutions combined (i.e., when dancing together) reveal candidate eliminations that are impossible to see when each of the solutions dances alone. The following joint-elimination tricks are useful:

> a. *Eliminate unmarked candidates*. If a cell contains a square (or under-dots) and also a circle (or over-dots), then any numbers in the cell that are unmarked can be erased because they cannot be part of either solution. These erasures should be followed by Step 3 cleanups of the whole puzzle, and more erasures, focusing on the structures that contain modifications.

---

[8] In the case starter cells, better results are obtained with cells whose structures also contain numerous cells with only two candidates.
[9] You can also use all the tricks of Step 1 (e.g., footnote 2) to quickly identify hidden singles in the squares.

[Special case: If a circle and a square happen to enclose the same number; then this number is a single and can be written in ink.]

b. *Eliminate double-blocked candidates*. If a structure contains a number enclosed in a circle, and the same number is enclosed in a square in a second structure that overlaps with the first, then all other instances of the number can be erased from all the cells in the intersection of the two structures.[10] This is a double-blocking. These erasures should be followed by Step 3 cleanups, and additional erasures. [Special case: When the two structures coincide, i.e., the circle and the square containing the same number are in the same structure then all other instances of the number can be erased from the structure.]

These two tricks eliminate candidates before you know which of the dancing solutions is right. This is helpful because sometimes identifying the correct solution is hard.

Several things can happen as you proceed thanks to the joint-elimination tricks. If you are lucky, one of the potential solutions may be developed to the end without a conflict and solve the puzzle. If you are a little less lucky, one of the potential solutions may lead to a conflict (violation of Sudoku Sub-rule 3a); or to an inability to complete a structure (violation of Sub-rule 3b). Then you know that this potential solution is wrong, and that the part of the other solution that you have identified is right. Therefore, you should then do four things: (i) erase all the markings of the wrong solution; (ii) write in ink all the identified singles corresponding to the retained solution, i.e., those numbers enclosed by the geometric figure (circle or square) corresponding to said solution; (iii) in cells that have markings for the retained solution (over-dots or under-dots), erase all candidates that do not have the marking of the retained solution; and (iv) do a Step 3 clean-up of what remains. Frequently, this will yield the final solution.

When the procedure does not lead to a solution (this can occur whether you have eliminated many candidates or none) you should erase all the markings and repeat the minuet (Step 4) using another starter item (either a cell or a half double). In some very rare cases, you may have to use four and even more starter items before the puzzle yields, but the author has yet to find a puzzle that cannot be solved after a few minuets.

## 4. CONCLUSION

This paper has introduced an algorithm for humans to use to solve 9×9 Sudoku puzzles with a unique solution. The method consists of a well-defined sequence of steps that can be translated into a computer algorithm. After hundreds of tries by hand, the author has yet to find a well-posed puzzle that cannot be solved. Therefore the following is proposed.

**The Minuet Conjecture**: *The minuet method solves all well-posed* 9×9 *Sudoku puzzles*.  □

---

[10] Otherwise both dancing solutions would have a conflict. Neither could be valid and the puzzle would not have a solution.

Further research is required to determine if the conjecture is true. Perhaps the algorithm should be programmed for computer application in the hunt for counterexamples. However, regardless of the outcome, the author's experience establishes on statistical grounds that the fraction of "evil" Sudoku puzzles that cannot be solved by the minuet method must be rather small—given the hundreds of puzzles successfully solved so far, it is estimated to be less than 1% with 90% confidence.

**Acknowledgement:** My friends B.J. Miller and Gil Rodriguez provided numerous suggestions and editorial comments. Their insights are gratefully acknowledged.